 \documentclass[superscriptaddress,aip,reprint,twocolumn]{revtex4-1}

\usepackage{graphicx,epsfig}
\usepackage{dcolumn}
\usepackage{bm}
\usepackage{textcomp}
\usepackage{amsmath}
\usepackage{amssymb}
\usepackage{gensymb}
\usepackage{color}
\usepackage{soul}

\usepackage[]{fontenc}
\draft 
\newcommand{\lyxmathsym}[1]{\ifmmode\begingroup\def\b@ld{bold}
  \text{\ifx\math@version\b@ld\bfseries\fi#1}\endgroup\else#1\fi}

\newcommand{\LA}{$\mathrm{LuAl_3}$}
\newcommand{\YA}{$\mathrm{YbAl_3}$}

\begin{document}


\title{Epitaxial growth and electronic properties of mixed valence YbAl$_3$ thin films} 


\affiliation{Laboratory of Atomic and Solid State Physics, Department of Physics, Cornell University, Ithaca, New York 14853, USA}
\affiliation{School of Applied and Engineering Physics, Cornell University, Ithaca, New York 14853, USA}
\affiliation{School of Electrical and Computer Engineering, Cornell University, Ithaca, New York 14853, USA}
\affiliation{Department of Materials Science and Engineering, Cornell University, Ithaca, New York 14853, USA}
\affiliation{Kavli Institute at Cornell for Nanoscale Science, Ithaca, New York 14853, USA}

\author{Shouvik Chatterjee}
\affiliation{Laboratory of Atomic and Solid State Physics, Department of Physics, Cornell University, Ithaca, New York 14853, USA}
\author{Suk Hyun Sung}
\affiliation{School of Applied and Engineering Physics, Cornell University, Ithaca, New York 14853, USA}
\author{David J. Baek}
\affiliation{School of Electrical and Computer Engineering, Cornell University, Ithaca, New York 14853, USA}
\author{Lena F. Kourkoutis}
\affiliation{School of Applied and Engineering Physics, Cornell University, Ithaca, New York 14853, USA}
\affiliation{Kavli Institute at Cornell for Nanoscale Science, Ithaca, New York 14853, USA}
\author{Darrell G. Schlom}
\affiliation{Department of Materials Science and Engineering, Cornell University, Ithaca, New York 14853, USA}
\affiliation{Kavli Institute at Cornell for Nanoscale Science, Ithaca, New York 14853, USA}
\author{Kyle M. Shen}
\affiliation{Laboratory of Atomic and Solid State Physics, Department of Physics, Cornell University, Ithaca, New York 14853, USA}
\affiliation{Kavli Institute at Cornell for Nanoscale Science, Ithaca, New York 14853, USA}


\begin{abstract}
We report the growth of thin films of the mixed valence compound YbAl$_3$ on MgO using molecular-beam epitaxy. Employing an aluminum buffer layer, epitaxial (001) films can be grown with sub-nm surface roughness.  Using x-ray diffraction, \emph{in situ} low-energy electron diffraction and aberration-corrected scanning transmission electron microscopy we establish that the films are ordered in the bulk as well as at the surface. Our films show a coherence temperature of 37 K, comparable to that reported for bulk single crystals. Photoelectron spectroscopy reveals contributions from both \textit{f}$^{13}$ and \textit{f}$^{12}$ final states establishing that \YA\/ is a mixed valence compound and shows the presence of a Kondo Resonance peak near the Fermi-level.
\end{abstract}


\maketitle 

\section{Introduction}
Intermetallic compounds containing certain rare-earth elements with an open \emph{f}-shell, such as Yb, Ce, U, Eu, etc., in a periodic arrangement form the so-called Kondo lattice that exhibits a plethora of emergent exotic properties such as unconventional superconductivity,\cite{Allan:2013ek} quantum criticality,\cite{Yang:2012gq} hidden order,\cite{Chatterjee:2013gd} and non-Fermi liquid behavior.\cite{Kambe:2014if} Understanding how these emergent properties arise from competing energy scales, and the possibility to rationally manipulate these interactions, is essential for any prospect of exploiting these properties for potential applications. Growing these intermetallics in epitaxial thin film form opens up new avenues for dimensional confinement and strain engineering of the Kondo lattice,\cite{Shishido:2010im} investigation of  proximity effects,\cite{Wang:2013fp} and the creation of tunable electronic states via artificial superlattices.\cite{Shimozawa:2014ii}\\

In intermetallic \YA\/, the Yb valence fluctuates between two configurations, Yb$^{2+}$(\textit{f}$^{13}$) and Yb$^{3+}$(\textit{f}$^{12}$).\cite{Kumar:2008es} Its ground state is non-magnetic and has a Kondo temperature of T$_{K}$ $\approx$ 670 K. Transport and thermodynamic measurements indicate that there exists another low temperature scale, T* $\approx$ 34K - 40K, below which Fermi-Liquid-like behavior has been found to emerge.\cite{Cornelius:2002ju} Moreover, it undergoes a much slower crossover from the high temperature local moment regime to the low temperature Fermi liquid regime than predicted by the Single Impurity Anderson Model (SIAM).\cite{Cornelius:2002ju,Bauer:2004fk} These observations, along with the anomalies in the temperature-dependent soft x-ray photoemission spectroscopy measurements of bulk and Lu-doped single crystals, suggest that Kondo lattice effects play an important role.\cite{Suga:2005dv,Yamaguchi:2007fs} The microscopic mechanism governing this behavior, however, remains enigmatic. One reason is that \YA\/ does not naturally cleave, which has precluded measurements of its electronic structure by momentum-resolved techniques such as angle resolved photoemission spectroscopy (ARPES) and spectroscopic imaging scanning tunneling microscopy (SI-STM). Measurements of fractured or scraped single crystals are complicated by multiple crystallographic faces\cite{Wahl:2011cm} and  issues with surface quality, coupled with a propensity of the \YA\/ surface to form oxides.\cite{Blyth:1993vb} One way of circumventing this difficulty is to synthesize epitaxial thin films and to measure them \emph{in situ} with spectroscopic probes. Here, we report the first progress along this direction.\\

\section{Experimental Details}
Thin films were grown on MgO substrates in a Veeco GEN10 MBE system with a liquid nitrogen cooled cryoshroud with a base pressure below 2$\times$10$^{-9}$ torr.  Prior to growth, MgO substrates were annealed in vacuum for 20 minutes at 800$\degree$C. Epitaxial (001) \YA\/ films were achieved by first depositing a 1 - 2 nm thick aluminum buffer layer, followed by a \LA\/ buffer layer on which \YA\/ was grown. For the deposition of the \LA\/ and \YA\/, elements were co-deposited onto a rotating substrate from Langmuir effusion cells at a growth rate of $\approx$ 0.4 nm/min. During growth \emph{in situ} reflection high-energy electron diffraction (RHEED) was used to monitor the surface evolution. After growth, the films were immediately transferred under ultra-high vacuum (5$\times$10$^{-10}$ torr) into an analysis chamber where low-energy electron diffraction (LEED) and \emph{in situ} photoemission spectroscopy were performed. Bulk structural characterization was performed \emph{ex situ} by four-circle x-ray diffraction (XRD) using Cu K$_{\alpha}$ radiation. The atomic scale structure of the \YA\/ thin film was studied by aberration-corrected scanning transmission electron microscopy (STEM) and electron energy loss spectroscopy (EELS) performed on a FEI Titan Themis 300. Cross-sectional TEM specimen were prepared using a FEI Strata 400 Focussed Ion Beam with a final milling step of 2 KeV to reduce surface damage. Films were further characterized by atomic force microscopy (AFM), scanning electron microscopy (SEM) and resistivity measurements. \\

\section{Results and Discussion}
\begin{figure*}[t]
\includegraphics[width=1\textwidth]{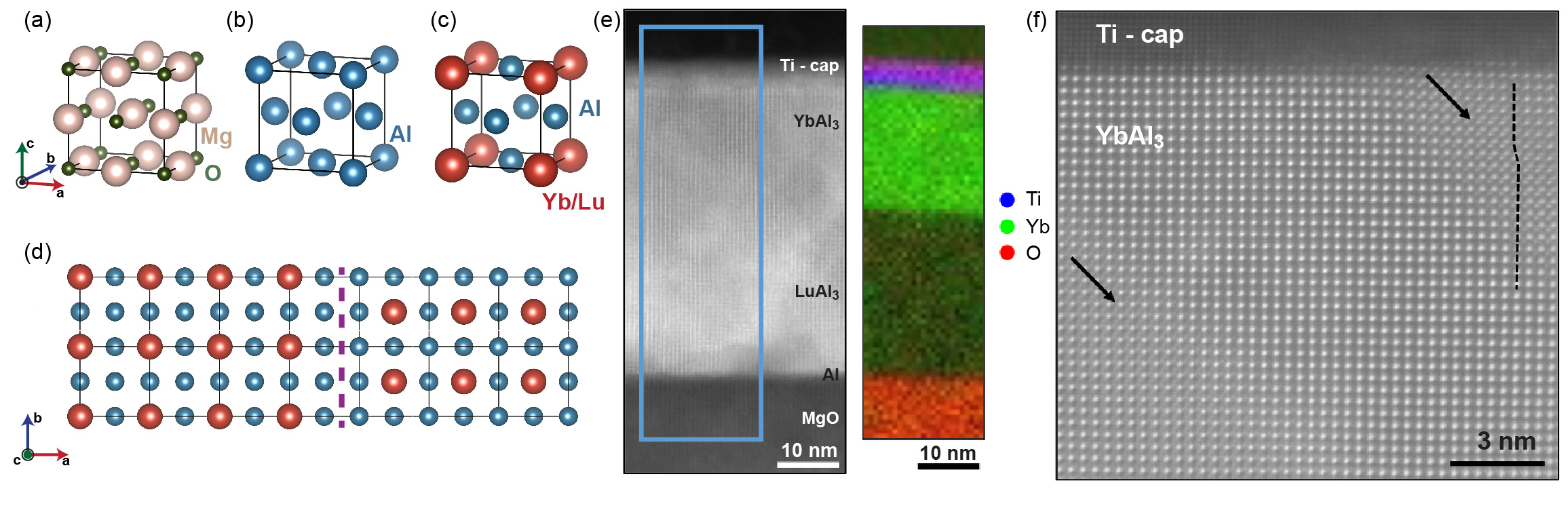}
\caption{Crystal structure of (a) MgO, (b) Al, and (c) \YA\//\LA\/. (d) Illustration of two possible domains of \YA\//\LA\/ on Al  when viewed along the [001] direction  and an anti-phase domain boundary (violet dotted line) between them. (e) High resolution high angle annular dark field scanning transmission electron microscopy (HAADF-STEM) image and  elemental map of a 20 nm thick \YA\/ film grown on 30nm  thick \LA\/ and 1.2 nm Al buffer layers. The film is capped by 6 nm of Ti to protect the film from oxidation. EELS mapping was performed in the area marked in blue (f) HAADF-STEM image showing the atomic arrangement of Yb atoms in  the \YA\/ film. Anti-phase domain boundaries are highlighted with black arrows. Dotted black line shows the shift of Yb atomic positions by ($\frac{1}{2}$, $\frac{1}{2}$, 0) across an anti-phase domain boundary.}\label{TEM}
\end{figure*}

 Both \YA\/ and  its structural analogue \LA\/ crystallize in a cubic AuCu$_{3}$ (L12)  structure where Yb/Lu atoms occupy the vertices of the unit cell and Al atoms occupy the face-centered positions (Fig. \ref{TEM}(c)). Various compounds belonging to the same space group, such as AuCu$_{3}$ have been reported to show an order-disorder transition, where above a certain temperature the disordered face-centered cubic (FCC) phase is stabilized.\cite{Cullity} In the disordered FCC phase occupation probability of all the lattice sites becomes $\frac{1}{4}$ for Yb/Lu and $\frac{3}{4}$ for Al atoms.  Diffraction measurements can distinguish the cubic ordered phase, as its diffraction peaks should be observed for all values of $h,k,\ell$ indices, while for the disordered phase, the structure factor becomes zero when indices are mixed between odd and even $h,k,\ell$ indices.  \YA\/ and \LA\/ have lattice constants of \textit{a} = 4.2 \AA\/  and \textit{a} = 4.19 \AA\/, respectively. The lattice constants of MgO and aluminum are 4.21 \AA\/  and 4.05 \AA\/, respectively, which provide a good lattice match to the \LA\/ or \YA\/ film. MgO and Al have a different space group ($Fm$$\overline{3}m$) than that of ordered \YA\//\LA\/ ($Pm$$\overline{3}m$), but the surface atomic arrangement of the oxygen atoms on the (001) surface of MgO and the Al atoms on the Aluminum (001) surface provide an excellent template for the epitaxial integration of \YA\//\LA\/ on MgO substrates with aluminum as a buffer layer (Figs. \ref{TEM}(a)-(c)). The face-centered sites and vertices are inequivalent for ($Pm$$\overline{3}m$) and thus, the \LA\//\YA\/ layers can nucleate either where the \LA\//\YA\/ is in perfect registry with that of the Al unit cell (left domain in Fig. \ref{TEM}(d)) or where it is shifted by ($\frac{1}{2}$, $\frac{1}{2}$, 0) (right domain in Fig. \ref{TEM}(d))  leading to the formation of anti-phase domains (Fig. \ref{TEM}(d)). Our EELS map confirm that there is negligible inter-diffusion across the boundary between the \YA\/ and \LA\/  layers (Fig. \ref{TEM}(e)). \\
 
\begin{figure}[t]
\includegraphics[width=1\columnwidth]{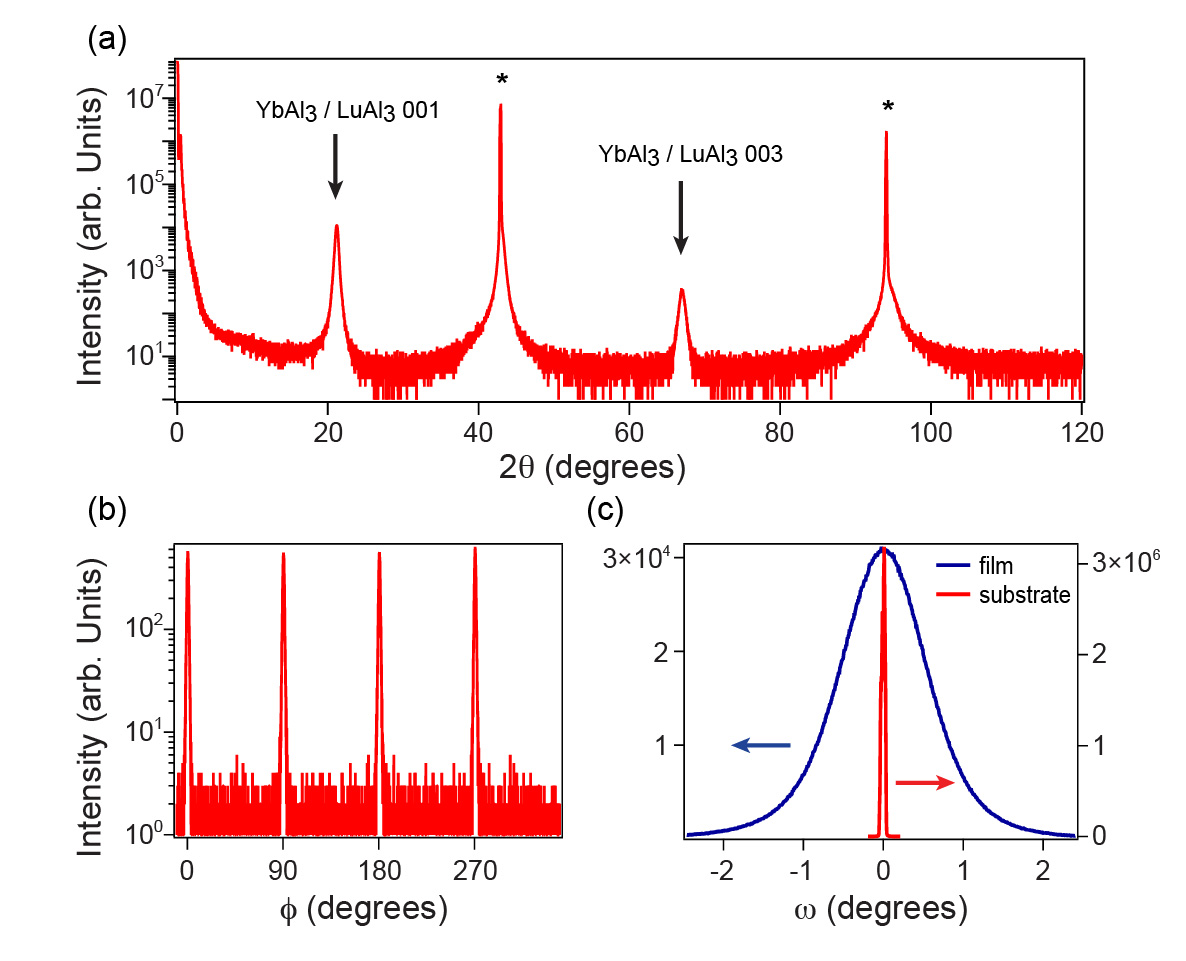}
\caption{(a) Out-of-plane $\theta$-2$\theta$ scan of a film with identical composition as in Fig. \ref{TEM}(e) and (f) but without a Ti capping layer. 001 and 003 peaks of the ordered L12 structure (shown in Fig. 1 (c)) establish that the film is ordered. Substrate peaks are marked by asterisks. (b) Azimuthal $\phi$ scan of the 101 \YA\/ diffraction peak at $\chi$ = 45$\degree$ , where $\chi$ = 0$\degree$  aligns the diffraction vector perpendicular to the plane of the substrate. $\phi$ = 0$\degree$ corresponds to the in-plane component of the diffraction vector aligned parallel to the [100] direction of the (001) MgO substrate. The $\phi$ scan in (b) establishes the epitaxial relationship to be (001)[100] \YA\/ $||$ (001)[100] MgO. (c) Rocking curve comparison between the 001 film peak and 002 substrate peak.} \label{XRD}
\end{figure}

 In Fig. \ref{XRD}(a), we show an out-of-plane $\theta$-2$\theta$ XRD scan. Prominent 001 and 003 Bragg peaks can be easily identified, confirming the films have the ordered L12 structure.\cite{YAorder} Moreover, the observation of only 00$\ell$ film peaks establishes that our films are single phase with the desired (001) \YA\/ out-of-plane orientation. An analysis of the peak positions reveal that the \LA\/ layer is relaxed while \YA\/ is strained to the underlying \LA\/ layer. An azimuthal $\phi$ scan of the (110) \YA\/ film peak is shown in Fig. \ref{XRD}(b). The $\phi$ scan in combination with the $\theta$ - 2$\theta$ scans establishes the epitaxial relationship of \YA\//\LA\/ with respect to MgO where (001)[100]  \YA\//\LA\/ $|| $(001)[100] MgO. Rocking curve measurements were performed on the 001 film peak and the 002 substrate peak, shown in Fig.\ref{XRD} (c). The full width at half maximum (FWHM) of the the film peak was found to be 0.32 degrees while that for the substrate peak was 0.03 degrees.

\begin{figure*}
\includegraphics[width=1\textwidth]{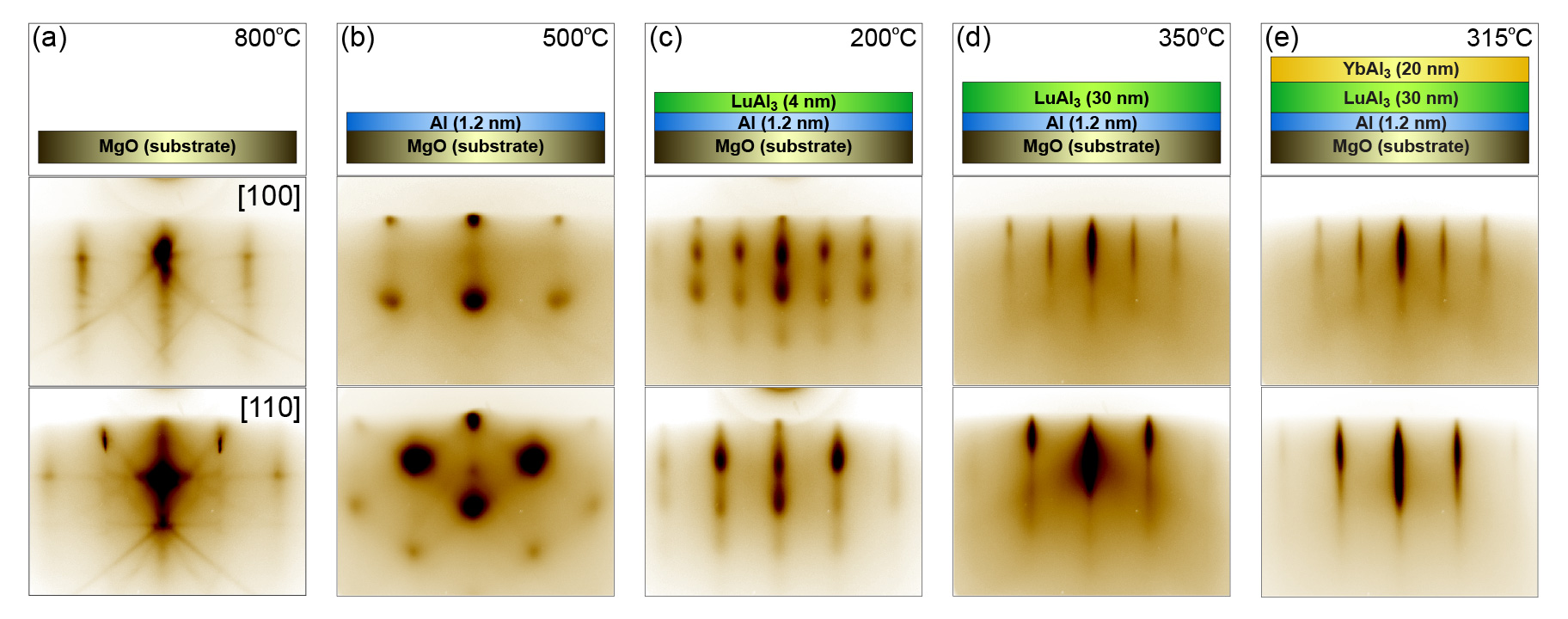}
\caption{RHEED images at different stages of the growth of a \YA\//\LA\//Al/MgO heterostructure. (a)-(e) Schematics of the layers present in the thin film heterostructure at each stage of growth at which RHEED patterns were recorded. Substrate temperatures are noted in the top right corner of the top panel. Corresponding RHEED spectra along the [100] and [110] azimuths are in the middle and the bottom panels, respectively.} \label{RHEED}
\end{figure*}

The growth process was initiated first by depositing an aluminum buffer layer at a substrate temperature of 500$\degree$C. RHEED patterns taken after the deposition process shown in Fig. \ref{RHEED}(b) reveal large transmission spots indicative of 3D grains.  Although depositing aluminum at elevated temperatures results in a rough surface, it also nucleates epitaxially oriented grains providing a template for the growth of epitaxially oriented \YA\//\LA\/ layers. Growing \YA\//\LA\/ directly on MgO or depositing the Al buffer layer at a lower temperature resulted in a mixed orientation where a secondary orientation with (111) \YA\//\LA\/ $||$ (001) MgO was observed. Films grown directly on MgO were also found to be much rougher by AFM, as \YA\//\LA\/ layers do not wet the MgO (001) surface well. Subsequently, a second buffer layer of \LA\/ was deposited at a substrate temperature of 200$\degree$C and annealed in vacuum at 350$\degree$C for 30 minutes. The deposition of the \LA\/ buffer layer was also important to achieve epitaxial \YA\/ films, as growing \YA\/ directly on the aluminum buffer layer resulted in secondary phases. As the \LA\/ growth proceeded, the RHEED pattern became more and more two-dimensional as individual grains coalesced into a flat surface (Fig. \ref{RHEED}(d)). Thicker \LA\/ buffer layers and post-annealing resulted in better quality films. Finally, the sample was cooled down to 200$\degree$C for the deposition of the \YA\/ layer. The growth temperature was ramped up to 315$\degree$C after initiating growth at 200$\degree$C. During growth, half-order peaks can be seen in RHEED images taken along the [100] azimuth (Fig. \ref{RHEED}(e)), indicating that the \YA\/ film grows with the ordered L12 structure (Fig. \ref{TEM}(c)). \\

\begin{figure}
\includegraphics[width=1\columnwidth]{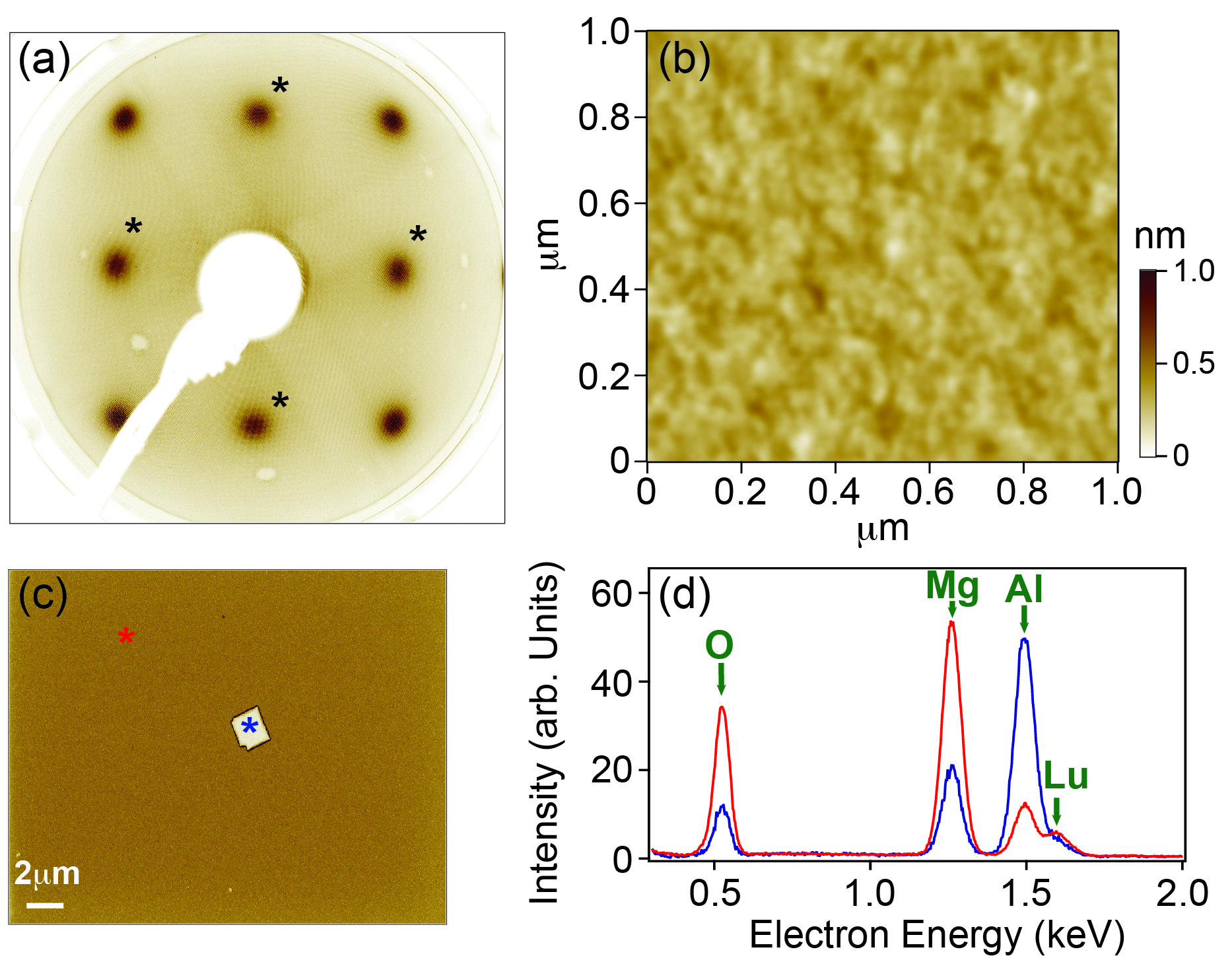}
\caption{(a) LEED image taken on a 20 nm thick uncapped \YA\/ film. Diffraction peaks with mixed indices are marked by asterisks. (b) AFM image of the same film. The measured rms surface roughness is $\approx$ 0.18 nm (c) SEM image of a 35 nm thick \LA\/ film corroborating the smoothness of the films over a large area. (d) EDX scans taken in regions marked by asterisks in (c).} \label{AFM}
\end{figure}

\emph{In situ} LEED patterns (Fig. \ref{AFM}(a)) indicate that the surface of the films is ordered with no additional surface reconstruction. The smoothness of the films was confirmed by AFM measurements (Fig. \ref{AFM}(b)) revealing a smooth surface with an rms roughness of only $\approx$ 1.8 \AA\/ over a 1 $\mu$m by 1 $\mu$m field of view.  SEM and corresponding energy dispersive x-ray spectroscopy (EDX) measurements on a 35 nm thick \LA\/ film with a 1.2 nm aluminium buffer shown in (Figs. \ref{AFM}(c)-(d)) reveal small aluminium-rich regions where excess aluminium has precipitated out. This observation is in agreement with the Yb-Al phase diagram (at temperatures relevant to our study a two-phase Al + \YA\/ region exists).\cite{phase} The expulsion of excess aluminium from the film ensures robustness of the growth method against slight flux mismatch, ensuring correct stoichiometry of our films.\\
 
\begin{figure}
\includegraphics[width=1\columnwidth]{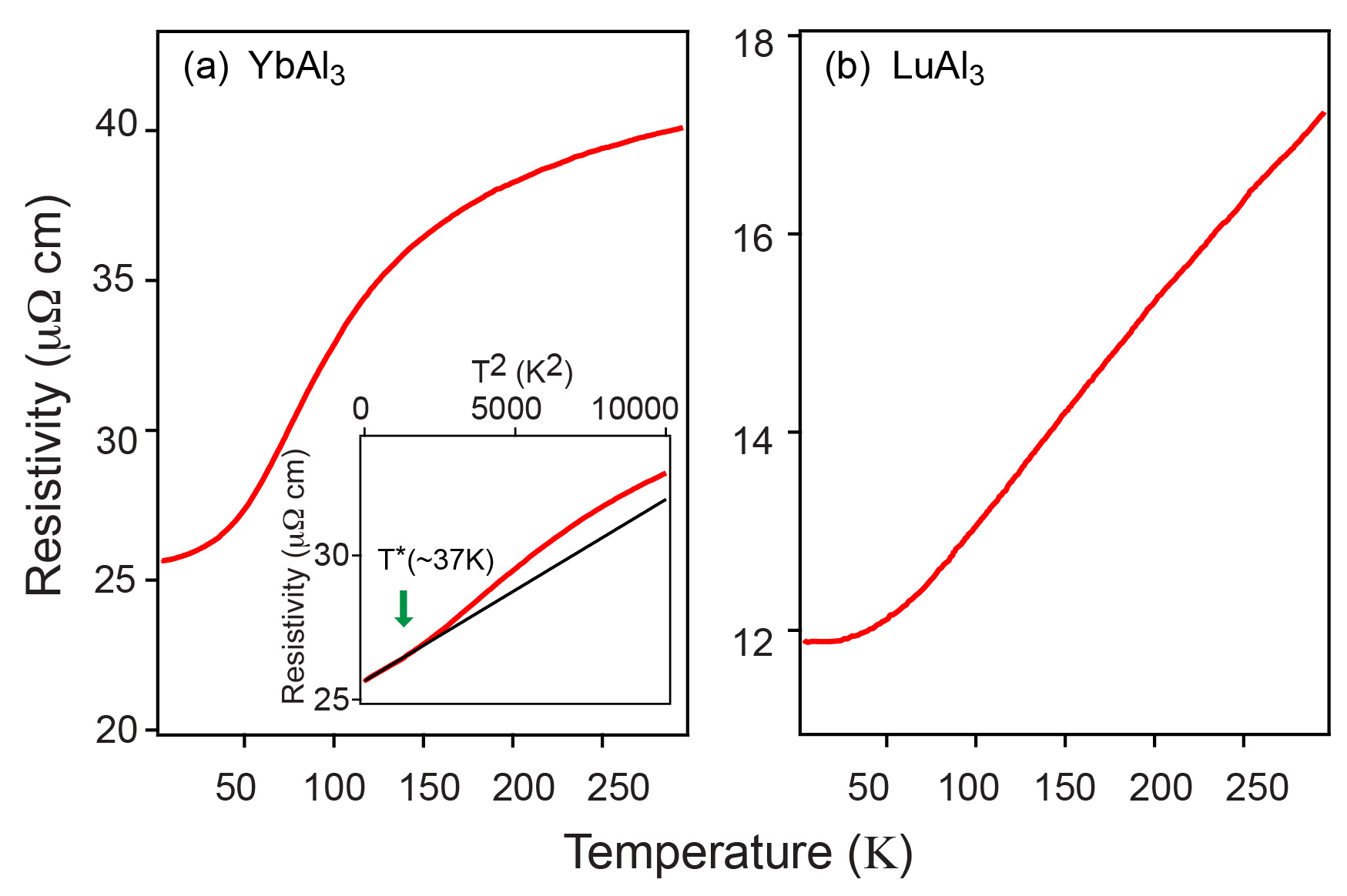}
\caption{Temperature-dependent resistivity of (a) 20 nm \YA\// 1.6 nm \LA\// 1.2 nm Al and of (b) 35 nm \LA\// 1.2 nm Al. Inset in (a) shows the onset of coherence at $\approx$ 37 K in the 20 nm thick \YA\/ film above which the resistivity starts to deviate from T$^{2}$ behavior. (black line)} \label{resistivity}
\end{figure}

Having investigated the structural aspects of our thin films, we now turn to their electronic properties. Four-point resistivity measurements were performed in a van der Pauw geometry using both a home-built dipper setup and a Quantum Design PPMS system. The temperature-dependent resistivity for a 20 nm \YA\// 1.6 nm \LA\// 1.2 nm Al and 35 nm \LA\// 1.2 nm Al thin films are shown in Figs. \ref{resistivity}(a) and \ref{resistivity}(b), respectively.  The resistivity curves are found to be qualitatively similar to that obtained for corresponding single crystals.\cite{Ohara:2001vg} The resistivity of \YA\/ is plotted as a function of T$^{2}$ in the inset of Fig. \ref{resistivity}(a), which shows that the film deviates from T$^{2}$ Fermi liquid behavior ($\rho(T) = \rho_{0} + AT^{2} $) above T$^{*}$ $\approx$ 37 K, similar to single crystals,\cite{Cornelius:2002ju} albeit with a lower residual resistivity for the single crystals. The roughness of the aluminum buffer layer employed at the beginning of the growth process and the presence of anti-phase domain boundaries could be potential contributors to both the higher residual resistivity and wide rocking curves (Fig. \ref {XRD}(c)) of our thin films. The T$^{2}$ coefficient of resistivity in our thin films is A = 6$\times$10$^{-4}$ $\pm$ 1$\times$10$^{-4}$ $\mu\Omega$ cm/K$^{2}$ similar to the earlier reported values for single crystals.\cite{Maple:1992, Ebihara:2000} Assuming a similar value of $\gamma$ = 45 mJ/mol K$^{2}$ in our thin films as in single crystals, we obtain a Kadowaki Woods ratio (A/$\gamma^{2}$) of $\approx$ 3$\times$10$^{-7}$ $\mu\Omega$ cm mol$^{2}$ K$^{2}$/mJ$^{2}$, which is orders of magnitude lower than the universal behavior observed in many $\textit{f}$ electron systems, A/$\gamma^{2}$ =  1$\times$10$^{-5}$ $\mu\Omega$ cm mol$^{2}$ K$^{2}$/mJ$^{2}$.\cite{Woods:1986} This apparent discrepancy can be reconciled by incorporating corrections due to the large degeneracy (N = 8) of the Yb 4$\textit{f}$ \textit{J} = 7/2 in the ground state that gives a value of A/$\gamma^{2}$ = 3.6$\times$10$^{-7}$ $\mu\Omega$ cm mol$^{2}$ K$^{2}$/mJ$^{2}$. \cite{Tsujii:2004} \\ 

\begin{figure}
\includegraphics[width=1\columnwidth]{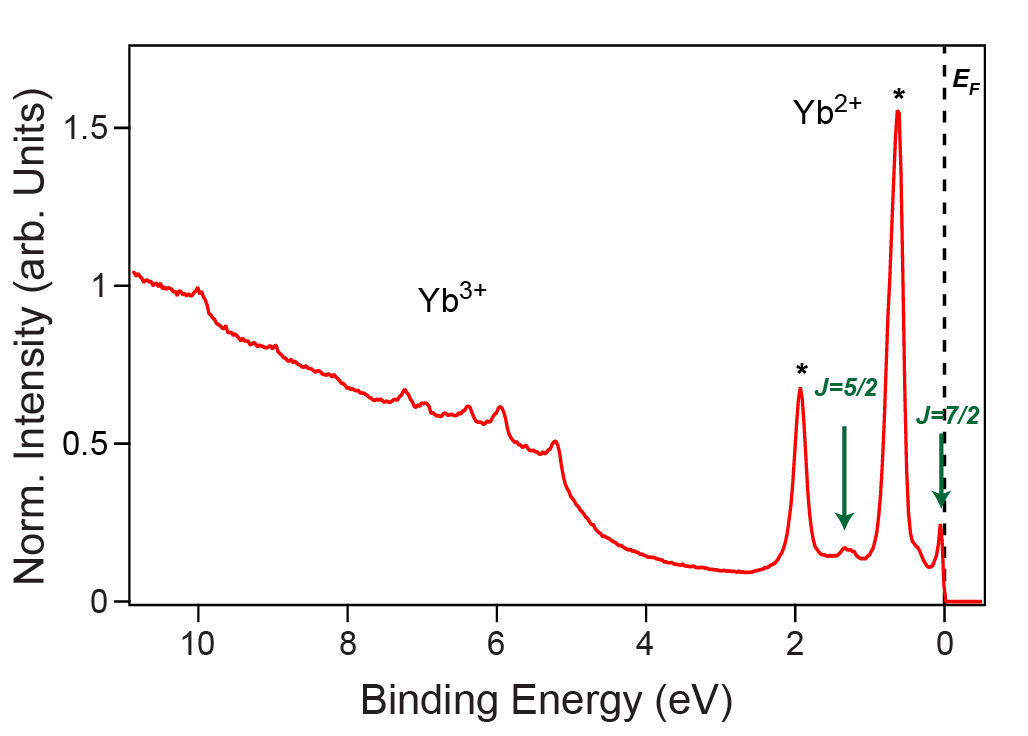}
\caption{PES spectra of \YA\/ obtained at 21 K. Peaks between 5 eV and 11 eV binding energy are derived from the Yb$^{3+}$\textit{f}$^{12}$ final state while those between 0 eV and 3 eV binding energy are derived from the Yb$^{2+}$\textit{f}$^{13}$ final state. The \textit{J} = 7/2 and \textit{J} = 5/2 spin orbit split states are indicated by arrows, whereas peaks marked with asterisks are the corresponding surface core level shifts.} \label{PES}
\end{figure}

Finally, Fig. \ref{PES}  shows \emph{in situ} photoemission spectra (PES) at normal emission obtained from these films. All measurements were taken using a monochromatized VUV5000 helium discharge lamp using a photon energy of 21.2 eV and a Scienta R4000 analyzer in a UHV chamber with base pressure better than 5$\times$10$^{-11}$ torr. The observed spectra show two sets of characteristic features corresponding to \textit{f}$^{13}$ and \textit{f}$^{12}$ final states, respectively, indicative of \YA\/ being a mixed valence compound. This is again consistent with earlier measurements of \YA\/ single crystals.\cite{Suga:2005dv,Tjeng:1993} Spectral peaks very close to the Fermi level with a binding energy of 0.03 eV and 1.34 eV, respectively, constitute features derived from Yb$^{2+}$\textit{f}$^{13}$  final states, where the former is derived from the Yb$^{2+}$\textit{f}$^{13}$ \textit{J} = 7/2  final state, also known as the Kondo resonance peak, while the latter is the spin orbit split peak derived from the Yb$^{2+}$\textit{f}$^{13}$ \textit{J} = 5/2  final state. Peaks marked with asterisks in Fig. \ref{PES} at binding energies 0.6 eV and 1.9 eV are corresponding surface core levels. The difference in binding energies between the bulk and surface doublets in our measurements is found to be $\approx$ 0.6 eV, consistent with earlier PES measurements.\cite{Suga:2005dv,Tjeng:1993} Spectral features observed at binding energies between 5 and 11 eV are derived from the Yb$^{3+}$\textit{f}$^{12}$ final states. Estimating the intensity of the photoemission peaks derived from Yb$^{2+}$ and Yb$^{3+}$ final states we obtain, in our thin films a mean Yb valence of 2.78 $\pm$ 0.06 at 21 K. The Yb valence estimated from our thin films is consistent with earlier low energy photoemission (2.77 at 10 K) \cite{Tjeng:1993} and x-ray spectroscopy measurements (2.78 at 20 K),\cite{Lawrence,Moreschini} but is higher than that estimated from soft x-ray PES (2.65 at 20 K).\cite{Suga:2005dv} Uncertainty in background subtraction leads to an uncertainty of $\approx$ 4$\%$ in our valence estimation, but our high energy resolution allows contributions form the surface states to be easily separated from that of the bulk states.

\section{Conclusion}
In conclusion, we have demonstrated that epitaxial thin films of \YA\/ (001) can be grown on MgO (001) by judicious choice of buffer layers and an optimized growth strategy. By providing access to pristine sample surfaces, our thin film approach could help elucidate microscopic mechanisms underlying the properties of \YA\/ by directly measuring the single particle spectral function in both real and momentum space via surface spectroscopic probes such as ARPES and STM. Furthermore, our approach opens the possibility to engineer the properties of \YA\/ and other mixed valence compounds via interfacial engineering or fabrication of heterostructures.\\

\section{Acknowledgements}
This work was supported by the National Science Foundation through DMR-0847385 and by the Cornell Center for Materials Research with funding from the NSF MRSEC program (DMR-1120296), and by the Research Corporation for Science Advancement (2002S). This work made use of the Cornell Center for Materials Research Shared Facilities which are supported through the NSF MRSEC program (DMR-1120296). The FEI Titan Themis 300 was acquired through NSF-MRI-1429155, with additional support from Cornell University, the Weill Institute, and the Kavli Institute at Cornell. S.C acknowledges helpful discussions with J.T. Heron and H.P. Nair.


\begin{thebibliography}{99}
\bibitem{YAorder} 001 diffraction peaks of \YA\/ and \LA\/ overlap, which makes it difficult to directly estimate the contribution from each layer. Intensity of the 001 diffraction peak of films with different  \LA\/ and \YA\/ layer thicknesses was measured to confirm that both \YA\/ and \LA\/ layers are ordered.
\end{thebibliography}
\end{document}